# SET-BASED REACHABILITY FOR LOW-THRUST SPACECRAFT IN TWO-BODY AND CISLUNAR DYNAMICAL SYSTEMS

Jinaykumar Patel[*] and Kamesh Subbarao[†]

This paper investigates the application of zonotope-based reachability analysis to low-thrust spacecraft in both two-body and cislunar environments. Reachable sets are generated under two-body and circular restricted three-body (CR3BP) dynamics using set-based methods that approximate nonlinear systems via Taylor expansions. A state-dependent coefficient (SDC) parameterization is also explored to represent nonlinear dynamics in a pseudo-linear form, enabling efficient matrix-based propagation of reachable sets. Applications include Earth–Mars transfer and cislunar scenarios such as L1 and L2 Halo orbits and Near Rectilinear Halo Orbits (NRHOs). The resulting reachable sets are used for safe trajectory generation and tracking, with comparisons drawn between model predictive control (MPC) and LQR-based station-keeping. The proposed approach provides a scalable framework for analyzing spacecraft behavior under complex dynamics and control constraints.

## INTRODUCTION

The growing interest in cislunar space, driven by plans for sustained lunar exploration, construction of lunar gateways, and expanding human activities beyond Earth's orbit, demands robust techniques for spacecraft trajectory planning and control. As missions within this domain inherently involve complex nonlinear dynamics and significant uncertainties, effective methods to guarantee safe and reliable spacecraft operations are crucial.

Reachability analysis, which involves computing the set of all possible states a spacecraft can attain from given initial conditions over a specified timeframe and under allowable control inputs, is an essential tool for addressing these mission-critical challenges. It provides crucial information for collision avoidance[1], obstacle avoidance[2], safe path planning, and robust guidance, navigation, and control under uncertainty. Specifically, in cislunar space, where trajectories must navigate highly sensitive multi-body gravitational fields and limited thrust capabilities, precise reachability analysis can significantly enhance mission safety and efficiency.

Several computational approaches have been developed for reachability analysis, each with its strengths and limitations. The most popular approach for low-thrust spacecraft, particularly in cislunar trajectories, involves random sampling and propagation of numerous trajectories. For example, one approach leverages indirect trajectory optimization by solving multiple optimal control problems with randomly sampled objectives, generating extremal trajectories that approximate the boundary of the reachable set[3]. Additionally, recent advancements demonstrate that extremal trajectory generation does not necessarily require solving optimal-control problems but can instead rely on direct ODE integration[4]. However, this method does not allow targeted sampling of extremal

---

[*]Graduate Student, Mechanical and Aerospace Engineering, The University of Texas at Arlington, Arlington, TX 76019
[†]Professor, Mechanical and Aerospace Engineering, The University of Texas at Arlington, Arlington, TX 76019

trajectories within specific subspaces, such as position or velocity spaces. Despite their effectiveness, these sampling-based approaches often require extensive trajectory propagation to yield accurate boundary approximations. Moreover, mathematical guarantees provided by these methods typically only apply to the convex hull of reachable sets, presenting challenges when capturing non-convex shapes accurately. Computational demands are also significantly high when employing realistic ephemeris models, further limiting their practicality. In contrast, recent work has introduced a set-based reachability approach that leverages state-transition tensors to efficiently propagate reachable sets under multi-body gravitational dynamics[5].

Near Rectilinear Halo Orbits (NRHOs), a subclass of halo orbits around the L1 and L2 Lagrange points in the Earth-Moon system, have emerged as promising candidates for future lunar gateway missions[6]. These orbits offer advantageous properties such as low-energy transfer paths, consistent Earth visibility, and eclipse avoidance, making them attractive for long-duration cislunar operations. However, due to gravitational perturbations, navigational uncertainties, and actuator constraints, maintaining a spacecraft on NRHOs requires continuous station-keeping. A variety of station-keeping strategies have been explored in the literature, including control methods based on dynamical systems theory (e.g., X-axis crossing techniques and Cauchy-Green Tensors)[7], linear-quadratic regulators (LQR), sliding mode control, and model predictive control (MPC) approaches[8].

This paper addresses a set propagation methodology, building upon our previous research in hypersonic atmospheric re-entry reachability analysis[9–11]. Specifically, reachability analysis can be leveraged for path planning or generating a safe reference trajectory that ensures collision avoidance and respects spacecraft thrust limitations. These reference solutions then serve as the foundation for subsequent station-keeping approaches, such as MPC, which can robustly track the trajectory while accommodating actuator constraints and uncertainties. We extend our earlier work to the domain of low-thrust spacecraft reachable set computations for both two-body and cislunar dynamics. In addition to the baseline reachable set methodology, we develop a novel approach that formulates the dynamics in a state-dependent coefficient (SDC) form, enabling structured matrix-based reachable set propagation while retaining compatibility with the original set-based framework. Furthermore, we evaluate the performance of MPC for station-keeping by comparing it against traditional LQR-based controllers. The resulting framework enables computationally efficient and reliable mission design, trajectory planning, and safety verification in complex gravitational environments.

**BACKGROUND**

**Reachability**

For time-invariant systems, a state $x_1$ is deemed reachable if there exists an input capable of transitioning the system state from $x(t_0)$ to $x_1$ within a finite duration, denoted by $T = t_f - t_0$. Given a predetermined state $x_1$, one can compute a distinct control history to effectuate this transition, provided that the system is inherently reachable. On the other hand, by employing the system's initial state $x(t_0)$, the governing dynamics, and the bounded control vector $u$, one can derive a set encompassing all feasible values of $x(t)$ at any specific time $t$. This is achieved by methodically sampling the control vector over the interval $[t_0, t]$. Such a set is conventionally referred to as the *Reachable Set*.

**Set Representations**

In the context of geometric analysis, understanding the mathematical nuances of set representations is paramount. This section presents a concise exploration into two pivotal set representations: polyhedral sets, and zonotopes.

**Polyhedral Set:** Polyhedral sets, often encountered in linear programming and optimization, emerge as the intersection of a limited number of halfspaces. For matrices $\boldsymbol{H} \in \mathbb{R}^{m \times n}$ and vectors $\boldsymbol{h} \in \mathbb{R}^m$, the polyhedral set in $\mathbb{R}^n$, symbolized as $\mathcal{H}$, is formally articulated as

$$\mathcal{H} = \boldsymbol{x} \in \mathbb{R}^n : \boldsymbol{H}\boldsymbol{x} \leq \boldsymbol{h} \tag{1}$$

**Zonotope:** Given a center $\boldsymbol{c} \in \mathbb{R}^n$ and a generator matrix $\boldsymbol{G} \in \mathbb{R}^{n \times p}$, a zonotope $\mathcal{Z} \subset \mathbb{R}^n$ is defined as

$$\mathcal{Z} = \boldsymbol{c} + \boldsymbol{G}\xi : |\xi|_\infty \leq 1 \tag{2}$$

Zonotopes possess the important closure property under Minkowski addition. Specifically, given two zonotopes $\mathcal{Z}_1 = \mathcal{Z}(\boldsymbol{c}_1, \boldsymbol{G}_1)$ and $\mathcal{Z}_2 = \mathcal{Z}(\boldsymbol{c}_2, \boldsymbol{G}_2)$, their Minkowski sum is itself a zonotope and can be computed as

$$\mathcal{Z}_1 \oplus \mathcal{Z}_2 = \mathcal{Z}(\boldsymbol{c}_1 + \boldsymbol{c}_2, [\boldsymbol{G}_1 \; \boldsymbol{G}_2]) \tag{3}$$

**PROBLEM STATEMENT**

For modeling of the inputs to the system, the equations of motion can be written as

$$\dot{\boldsymbol{x}}(t) = \boldsymbol{f}(\boldsymbol{x}(t), t) + \frac{T_{\max}}{m} \boldsymbol{B}\boldsymbol{u}(t) \tag{4}$$

where $T_{\max}$ represents the maximum thrust of the propulsion system, $m$ denotes the mass of the spacecraft, and $\boldsymbol{B} = \left[\boldsymbol{0}_{\frac{n}{2} \times \frac{n}{2}}, \boldsymbol{I}_{\frac{n}{2} \times \frac{n}{2}}\right]^T$. The variation in mass due to propellant consumption is neglected, as its effect is assumed negligible over the short propagation durations considered. The set of admissible inputs is defined as

$$\mathcal{U} = \{\boldsymbol{u} \mid \|\boldsymbol{u}\|_2 \leq 1\} \tag{5}$$

The objective is to compute the set of all states that the spacecraft can reach under the dynamics given in Eq. (4), starting from a known initial condition and subject to allowable control inputs and bounded disturbances. This reachable set provides key guarantees for constraint satisfaction, including position bounds, obstacle avoidance, and terminal capture within a desired region.

We consider two approaches for computing these reachable sets. The first is a zonotope-based method in which a Taylor expansion is applied directly to the nonlinear function $\boldsymbol{f}(\boldsymbol{x}, t)$, and reachable sets are propagated using linearized approximations[10]. The second approach reformulates the system into a SDC form. This representation facilitates a matrix-level Taylor expansion, enabling the propagation of reachable sets through the evolution of parameterized system matrices. Both methods utilize zonotopic over-approximations for computational efficiency and are adapted to handle the nonlinearities inherent in multi-body gravitational dynamics.

Additionally, we assess the utility of the computed reachable sets for station-keeping near periodic orbits such as NRHOs. The reachable sets are used to generate reference trajectories that are subsequently tracked using a MPC. To evaluate performance, we compare MPC-based station-keeping with traditional LQR strategies under uncertainty. This framework supports scalable trajectory planning, robust control design, and safe operation in complex orbital regimes.

## DYNAMICAL SYSTEMS

**Two-Body Equations of Motion**

In the first case study, we consider the heliocentric transfer phase of an Earth-to-Mars rendezvous problem, neglecting third-body perturbations and assuming zero hyperbolic excess velocity at departure. Let $\boldsymbol{r} \in \mathbb{R}^3$ and $\boldsymbol{v} \in \mathbb{R}^3$ denote the spacecraft's position and velocity vectors, respectively, and define the state vector as $\boldsymbol{x} = [\boldsymbol{r}^\top, \boldsymbol{v}^\top]^\top \in \mathbb{R}^6$.

The governing dynamics under the two-body gravitational model with low-thrust propulsion are given by

$$\dot{\boldsymbol{x}} = \underbrace{\begin{bmatrix} \boldsymbol{v} \\ -\frac{\mu}{\|\boldsymbol{r}\|^3} \boldsymbol{r} \end{bmatrix}}_{\boldsymbol{f}(\boldsymbol{x},t)} + \frac{T_{\max}}{m} \underbrace{\begin{bmatrix} \boldsymbol{0}_{3\times 3} \\ \boldsymbol{I}_{3\times 3} \end{bmatrix}}_{\boldsymbol{B}} \boldsymbol{u}(t) \qquad (6)$$

where $\mu$ is the gravitational parameter of the Sun, and $\boldsymbol{u}(t) \in \mathbb{R}^3$ is the normalized thrust control input satisfying $\|\boldsymbol{u}(t)\|_2 \leq 1$, which encapsulates both throttle $\delta \in [0,1]$ and thrust direction $\hat{\boldsymbol{a}} \in \mathbb{R}^3$, such that $\boldsymbol{u}(t) = \delta \hat{\boldsymbol{a}}$.

This formulation aligns with the general dynamics structure described in Eq. (4) of the problem statement, enabling direct application of the proposed reachable set computation frameworks to this two-body transfer scenario.

**Circular Restricted Three-Body Problem**

The Circular Restricted Three-Body Problem (CR3BP) provides a foundational model for analyzing spacecraft motion in the Earth-Moon system, where the spacecraft, assumed to have negligible mass, is influenced only by the gravitational attraction of the two primaries: Earth ($P_1$) and Moon ($P_2$), with normalized masses $1-\mu$ and $\mu$, respectively.

The problem is formulated in a rotating barycentric reference frame in which the Earth and Moon are fixed on the $x$-axis. The state vector is given by $\boldsymbol{x} = [x, y, z, \dot{x}, \dot{y}, \dot{z}]^\top$, comprising the position and velocity of the spacecraft relative to the Earth-Moon barycenter.

In normalized units, where the Earth-Moon distance, total mass, and mean motion are unity, the equations of motion for a low-thrust spacecraft under CR3BP dynamics are written as

$$\dot{\boldsymbol{x}} = \underbrace{\begin{bmatrix} \dot{\boldsymbol{r}} \\ \boldsymbol{f}(\boldsymbol{r}, \boldsymbol{v}) \end{bmatrix}}_{\boldsymbol{f}(\boldsymbol{x},t)} + \frac{T_{\max}}{m} \underbrace{\begin{bmatrix} \boldsymbol{0}_{3\times 3} \\ \boldsymbol{I}_{3\times 3} \end{bmatrix}}_{\boldsymbol{B}} \boldsymbol{u}(t) \qquad (7)$$

where $\boldsymbol{r} = [x, y, z]^\top$, $\boldsymbol{v} = [\dot{x}, \dot{y}, \dot{z}]^\top$, and the gravitational acceleration term is given by:

$$\boldsymbol{f}(\boldsymbol{r}, \boldsymbol{v}) = \begin{bmatrix} x + 2\dot{y} - \frac{(1-\mu)(x+\mu)}{r_1^3} - \frac{\mu(x+\mu-1)}{r_2^3} \\ y - 2\dot{x} - \frac{(1-\mu)y}{r_1^3} - \frac{\mu y}{r_2^3} \\ -\frac{(1-\mu)z}{r_1^3} - \frac{\mu z}{r_2^3} \end{bmatrix},$$

with $r_1 = \sqrt{(x+\mu)^2 + y^2 + z^2}$ and $r_2 = \sqrt{(x+\mu-1)^2 + y^2 + z^2}$. The input $\boldsymbol{u}(t) \in \mathbb{R}^3$ represents the normalized thrust vector, constrained such that $\|\boldsymbol{u}(t)\|_2 \leq 1$.

This form aligns with the general system dynamics introduced in the Problem Statement Eq. (4), thereby enabling consistent use of zonotope-based reachable set computation methods for low-thrust trajectories in the CR3BP model.

## METHOD 1: REACHABLE SET COMPUTATION VIA DIRECT NONLINEAR TAYLOR APPROXIMATION

This section considers general nonlinear systems with uncertain parameters that are constrained within certain limits. The reachability of these nonlinear systems in CORA is determined by abstracting the state-space [12]. The continuous-time dynamics are modeled as Eq.(4). To perform reachable set analysis, the continuous-time dynamics are discretized using a suitable integration method, yielding a discrete-time model

$$\boldsymbol{x}_{k+1} = \boldsymbol{g}(\boldsymbol{x}_k, \boldsymbol{u}_k) \tag{8}$$

where $\boldsymbol{g}$ represents the discrete-time version of the continuous function $f$.

To simplify notation, we define the joint state-input vector $\boldsymbol{z}_k^T = [\boldsymbol{x}_k^T, \boldsymbol{u}_k^T]$. The function $\boldsymbol{g}_i$ for the $i$th state component is approximated using a Taylor series expansion

$$\begin{aligned}(\boldsymbol{x}_{k+1})_i = \ & \boldsymbol{g}_i(\boldsymbol{x}_k, \boldsymbol{u}_k) = \boldsymbol{g}_i\left(\boldsymbol{z}_k^*\right) + \nabla_{\boldsymbol{z}_k}\left(\boldsymbol{g}_i\left(\boldsymbol{z}_k^*\right)\right)\bigg|_{\boldsymbol{z}_k=\boldsymbol{z}_k^*} (\boldsymbol{z}_k - \boldsymbol{z}_k^*) + \\ & \frac{1}{2}(\boldsymbol{z}_k - \boldsymbol{z}_k^*)^T \nabla_{\boldsymbol{z}_k}^2\left(\boldsymbol{g}_i\left(\boldsymbol{z}_k^*\right)\right)\bigg|_{\boldsymbol{\zeta}} (\boldsymbol{z}_k - \boldsymbol{z}_k^*) + \ldots \end{aligned} \tag{9}$$

where $\boldsymbol{\zeta} \in \{\boldsymbol{z}_k^* + \alpha(\boldsymbol{z}_k - \boldsymbol{z}_k^*) \mid \alpha \in [0,1]\}$.

The infinite Taylor series is over-approximated by using its first-order representation with a Lagrange remainder

$$(\boldsymbol{x}_{k+1})_i \in \underbrace{\boldsymbol{g}_i\left(\boldsymbol{z}_k^*\right) + \nabla_{\boldsymbol{z}_k}\left(\boldsymbol{g}_i\left(\boldsymbol{z}_k^*\right)\right)\bigg|_{\boldsymbol{z}_k=\boldsymbol{z}_k^*}(\boldsymbol{z}_k - \boldsymbol{z}_k^*)}_{1^{st} \text{ order Taylor series}} + \underbrace{\frac{1}{2}(\boldsymbol{z}_k - \boldsymbol{z}_k^*)^T \nabla_{\boldsymbol{z}_k}^2\left(\boldsymbol{g}_i\left(\boldsymbol{z}_k^*\right)\right)\bigg|_{\boldsymbol{\zeta}}(\boldsymbol{z}_k - \boldsymbol{z}_k^*)}_{\text{Lagrange remainder } (L_k)_i} \tag{10}$$

The reachable set for the linearized system ($\mathcal{R}^{lin}(t)$) is first computed. The reachable set due to the Lagrange remainder, $\mathcal{R}^{err}(t)$, is evaluated using the nonlinear bounds. The total reachable set is the Minkowski sum

$$\mathcal{R}(t) = \mathcal{R}^{lin}(t) \oplus \mathcal{R}^{err}(t) \tag{11}$$

**Computation of the Set of Linearization Errors**

The second-order partial derivatives for the $i$th state component are given by

$$\boldsymbol{J}_i(\boldsymbol{\zeta}) = \nabla_{\boldsymbol{z}_k}^2\left(\boldsymbol{g}_i\left(\boldsymbol{z}_k\right)\right)\bigg|_{\boldsymbol{\zeta}} \tag{12}$$

The Lagrange remainder is then

$$(L_k)_i = \frac{1}{2}(z_k - z_k^*)^T J_i(\zeta)(z_k - z_k^*), \qquad (13)$$
$$\zeta \in \{z_k^* + \alpha(z_k - z_k^*) \mid \alpha \in [0,1]\}$$

For a zonotope $\mathcal{Z} = (c, g^{(1)}, \ldots, g^{(e)})$, the Lagrange remainder is over-approximated as

$$|(L_k)_i| \subseteq [0, (l_k)_i] \qquad (14)$$

where

$$(l_k)_i = \frac{1}{2}\gamma^T \max(|J_i(\zeta)|)\gamma, \quad \gamma = |c - z_k^*| + \sum_i^e \left|g^{(i)}\right| \qquad (15)$$

**Iterative Reachable Set Computation**

The reachable set is computed for each time segment $t \in [(k-1)\Delta t, k\Delta t]$

$$\mathcal{R}([0, t_f]) = \bigcup_{k=1}^{t_f/\Delta t} \mathcal{R}([(k-1)\Delta t, k\Delta t]) \qquad (16)$$

If the Lagrange remainder exceeds a user-specified tolerance, the current zonotope is split along the direction with the largest contribution to the remainder, and the reachability computation is repeated for the split sets.

## METHOD 2: REACHABLE SET COMPUTATION VIA STATE-DEPENDENT COEFFICIENT PARAMETERIZATION

Consider discrete-time, nonlinear function $f_d(x_k)$. We make the following standing assumption on the nonlinear functions $f_d : \mathbb{R}^n \to \mathbb{R}^n$. The nonlinear functions can be put into corresponding pseudo-linear forms using the SDC parameterization as

$$f_d(x_k) = A(x_k)x_k \qquad (17)$$

where $A : \mathbb{R}^n \to \mathbb{R}^{n \times n}$ is nonlinear matrix-valued function.

To this end, we recall the following useful result.

**Proposition 1**[13]. *SDC parameterization of $f_d(x_k)$ as in Eq. (17) always exist for some $\mathcal{C}^{t-1}$ matrix-valued functions $A : \mathbb{R}^n \to \mathbb{R}^{n \times n}$. This property is satisfied by the following parameterization*

$$A(x_k) = \int_0^1 \left.\frac{\partial f_d(x_k)}{\partial x_k}\right|_{x_k = \lambda x_k} d\lambda \qquad (18)$$

where $\lambda$ is a dummy variable of integration.

Note that multiple SDC parameterizations of the form Eq. (17) are possible for $n > 1$ using mathematical factorization[14].

For the discrete-time nonlinear system once we have the SDC parameterization Eq. (17), the key idea is to replace the nonlinear mapping with a family of linear systems whose coefficients lie within suitable bounds. Specifically, by encapsulating the matrix-valued function $A(x_k)$ within a (potentially time-varying) polytopic set, one can leverage convex hull or zonotope-based techniques originally developed for linear or LPV systems. This enables systematic construction of tube enclosures that reflect the local variation of $A(x_k)$. Because each step uses an updated approximation of the Jacobian, the resulting reachable sets can be less conservative than global bounding methods that do not exploit local linear structures.

**Zonotope-Based Matrix Taylor Expansion**

To propagate the reachability set through the nonlinear system, we seek to approximate the state-dependent matrix $A(x_k)$ over the uncertainty in $x_k$. Suppose the current state $x_k$ lies within a zonotope $\mathcal{Z}_k = (c_k, G_k)$, where $c_k$ is the center and $G_k$ is the generator matrix. We perform a first-order matrix Taylor expansion of $A(x_k)$ around $c_k$

$$A(x_k) \approx A(c_k) + \sum_{i=1}^{n} \left.\frac{\partial A}{\partial x_i}\right|_{c_k} (x_k - c_k)_i \tag{19}$$

Since $x_k - c_k = G_k \xi_k$, where $\xi_k \in [-1, 1]^q$, we substitute

$$(x_k - c_k)_i = \sum_{j=1}^{q} g_k^{(i,j)} \xi_j \tag{20}$$

Therefore, the matrix-valued function $A(x_k)$ becomes

$$A(x_k) \in \left\{ A(c_k) + \sum_{j=1}^{q} \tilde{G}_A^{(j)} \xi_j : \xi_j \in [-1, 1] \right\} \tag{21}$$

where each $\tilde{G}_A^{(j)} = \sum_{i=1}^{n} \left.\frac{\partial A}{\partial x_i}\right|_{c_k} g_k^{(i,j)}$ serves as a matrix generator that captures how $A(\cdot)$ varies over the zonotope.

**Reachable Set Computation**

The next state is given by

$$x_{k+1} = A(x_k) x_k \tag{22}$$

We compute the reachable set $\mathcal{R}_{k+1}$ using the matrix zonotope form of $A(x_k)$ and the state zonotope $\mathcal{Z}_k$.

Define the linearized propagation as

$$\mathcal{R}_k^{\text{lin}} = A(c_k) \mathcal{Z}_k \tag{23}$$

and the matrix perturbation contributions as

$$\mathcal{R}_k^{\text{dev}} = \sum_{j=1}^{q} \tilde{G}_A^{(j)} \mathcal{Z}_k^{[j]} \tag{24}$$

where $\mathcal{Z}_k^{[j]}$ are projections or scaled versions of $\mathcal{Z}_k$ depending on each generator direction. These terms are efficiently computed using standard zonotope arithmetic.

### Total Reachable Set and Refinement Strategy

Combining the linear part, matrix variation, and Lagrange remainder, the total reachable set becomes

$$\mathcal{R}_{k+1} = \mathcal{R}_k^{\text{lin}} \oplus \mathcal{R}_k^{\text{dev}} \oplus \mathcal{R}_k^{\text{err}} \tag{25}$$

To control conservativeness, we monitor the contribution of $\mathcal{R}_k^{\text{err}}$. If it exceeds a user-defined threshold, we bisect the zonotope $\mathcal{Z}_k$ along the generator with the highest contribution to the error, and recursively repeat the reachability analysis over the resulting sub-zonotopes. This SDC-based reachability analysis thus leverages the structure of pseudo-linear models, providing a principled framework for over-approximating nonlinear dynamics using matrix-valued zonotopic expansions.

## MODEL PREDICTIVE CONTROL FORMULATION

### Prediction Model

The spacecraft's dynamics are discretized as

$$\boldsymbol{x}_{k+1} = \boldsymbol{f}_d(\boldsymbol{x}_k, \boldsymbol{u}_k) \tag{26}$$

where $\boldsymbol{x}_k \in \mathbb{R}^{n_x}$ is the state at time step $k$, $\boldsymbol{u}_k \in \mathbb{R}^{n_u}$ is the control input, and $\boldsymbol{f}_d(\cdot)$ represents the discrete-time CR3BP dynamics computed via Runge–Kutta integration.

### Finite-Horizon Optimal Control Problem

At each sampling instant, an optimal control sequence is computed by minimizing a quadratic cost over a finite prediction horizon of length $N$:

$$\begin{aligned}
\min_{U} \quad & \sum_{i=0}^{N-1} \left(\boldsymbol{x}_{k+i|k} - \boldsymbol{x}_{k+i}^{\text{ref}}\right)^\top \boldsymbol{Q} \left(\boldsymbol{x}_{k+i|k} - \boldsymbol{x}_{k+i}^{\text{ref}}\right) + \boldsymbol{u}_{k+i|k}^\top \boldsymbol{R} \boldsymbol{u}_{k+i|k} \\
& + \left(\boldsymbol{x}_{k+N|k} - \boldsymbol{x}_{k+N}^{\text{ref}}\right)^\top \boldsymbol{P} \left(\boldsymbol{x}_{k+N|k} - \boldsymbol{x}_{k+N}^{\text{ref}}\right) \\
\text{s.t.} \quad & \boldsymbol{x}_{k+i+1|k} = \boldsymbol{f}_d(\boldsymbol{x}_{k+i|k}, \boldsymbol{u}_{k+i|k}), \quad i = 0{:}N-1, \\
& \boldsymbol{u}_{k+i|k} \in \mathcal{U}, \quad i = 0{:}N-1, \\
& \boldsymbol{x}_{k+i|k} \in \mathcal{R}_{k+i}, \quad i = 1{:}N,
\end{aligned} \tag{27}$$

where $N$ is the prediction horizon, $\boldsymbol{Q} \in \mathbb{R}^{n_x \times n_x}$ and $\boldsymbol{R} \in \mathbb{R}^{n_u \times n_u}$ are symmetric positive definite weighting matrices for the state and control, respectively, and $\boldsymbol{P} \succ 0$ is the terminal penalty. The set $\mathcal{U}$ is a zonotopic set reflecting the spacecraft's bounded thrust capabilities:

$$\mathcal{U} = \{\boldsymbol{u} \mid \boldsymbol{u}_{\min} \leq \boldsymbol{u} \leq \boldsymbol{u}_{\max}\} \tag{28}$$

and can be further represented by a zonotope if needed.

### Reachable Set Constraints for Safety

To ensure safety and feasibility under operational uncertainties, reachable sets $\mathcal{R}_{k+i}$ are computed online via SDC-based methods and incorporated as tube constraints on the predicted state trajectory. Thus, the MPC guarantees that all planned future states remain within the computed safe sets under the modeled uncertainties.

**Solver Implementation**

The optimization problem is implemented using `CasADi`[15] and solved using IPOPT. At each step, the decision variables include the predicted control sequence $U = \{u_{k|k}, \ldots, u_{k+N-1|k}\}$. State and input constraints are enforced at each stage, and the optimization is warm-started with the previous solution for efficiency.

**Remark:** For performance benchmarking, a LQR is also implemented for trajectory tracking and compared against MPC in terms of accuracy and constraint satisfaction.

## SIMULATION RESULTS

All simulations were carried out using `MATLAB` R2022a on a computer equipped with 16 GB of RAM and an Intel Core i7 processor running at 2.10 GHz. Planetary ephemerides and initial conditions were retrieved using NASA's JPL HORIZONS online solar system data and ephemeris service.

**Control Input Representation Using Zonotopes**

The thrust vector $u \in \mathbb{R}^3$ is constrained by a norm bound $\|u\|_2 \leq T_{\max}$, corresponding to the spacecraft's maximum allowable thrust. Instead of representing this constraint as a Euclidean ball, which is computationally expensive to propagate in reachable set analysis, we approximate it using a *zonotope centered at the origin* with carefully selected generator vectors. In particular, we construct the thrust zonotope $\mathcal{Z}_u \subset \mathbb{R}^4$, where the first three components represent the thrust direction and the fourth dimension captures the total thrust magnitude (or throttle). The zonotope is defined as:

$$\mathcal{Z}_u = \left\{ c + \sum_{i=1}^{m} \beta_i g_i \,\middle|\, \beta_i \in [-1, 1] \right\}, \quad c = \begin{bmatrix} 0 \\ 0 \\ 0 \\ T_{\max} \end{bmatrix},$$

where each generator $g_i \in \mathbb{R}^4$ contributes a direction in the control space. A minimal set includes the axis-aligned directions $T_{\max} \cdot e_1$, $T_{\max} \cdot e_2$, and $T_{\max} \cdot e_3$ corresponding to thrust along the principal axes. To better approximate the norm-bounded thrust region, additional generators are added along diagonal directions (e.g., $[1, 1, 0, 0]^\top$, $[1, 0, 1, 0]^\top$) and scaled appropriately. This ensures the zonotope conservatively encloses the true thrust constraint set while remaining computationally efficient for propagation within reachability analysis frameworks.

Table 1: Earth-to-Mars reachable set propagation times.

| Time of Flight (days) | Computation Time (s) |
|---|---|
| 200 | 3.443 |
| 300 | 5.103 |

**Two-Body Dynamics Results**

The results obtained for the reachable set calculations for Earth-to-Mars transfer are presented in this section. This example is adapted from a previously published study[16]. For all results, $T_{\max} = 0.3$ N, $I_{sp} = 3000$ s, and $m_0 = 1000$ kg. The initial position and velocity vectors are $\boldsymbol{r} = [-140699693; -51614428; 980]^\top$ km and $\boldsymbol{v} = [9.774596; -28.07828; 4.337725\text{E-}4]^\top$ km/s, respectively. The complete state vector of the spacecraft at the beginning of the maneuver is given as

$$\boldsymbol{x} = [-140699693;\ -51614428;\ 980;\ 9.774596;\ -28.07828;\ 4.337725\text{E-}4]^\top \tag{29}$$

with units km and km/s corresponding to a departure epoch of 10 April 2007.

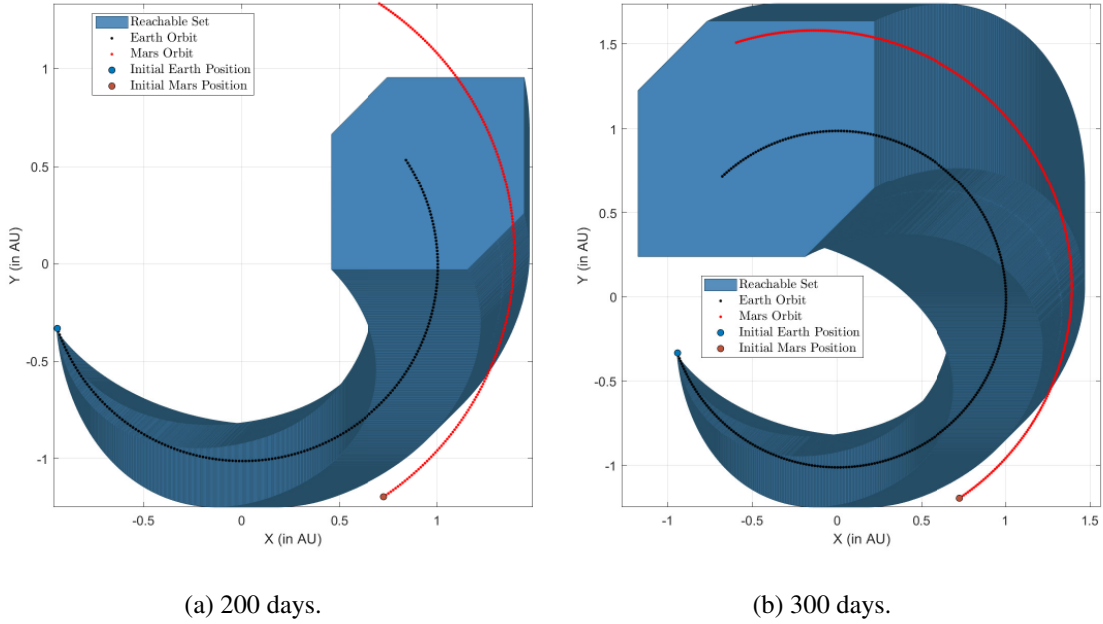

(a) 200 days.

(b) 300 days.

**Figure 1**: Depiction of the Earth-to-Mars position reachable set.

Figure 1 illustrates the growth of the reachable position set over increasing time horizons, specifically for flight durations of 200 and 300 days. The trajectory of Mars (shown in red) highlights the evolution of the target's position with increasing time of flight. It is evident from the figure that a minimum flight duration of over 200 days is required to achieve reachability. This positional reachability forms the first step in evaluating feasibility for a rendezvous maneuver. Data indicating the time of flight and simulation run times are presented in Table 1.

**Circular Restricted Three-Body Dynamics Results**

All quantities in the CR3BP simulation results are expressed in nondimensional units. The normalization is performed by defining a set of characteristic scaling factors. The characteristic length $l^*$ is chosen as the distance between the two primary bodies, Earth and Moon, and is set to $l^* = 3.844 \times 10^5$ km. The characteristic mass $m^*$ is defined as the combined mass of the primaries, given by $m^* = m_1 + m_2 = 6.0458 \times 10^{24}$ kg. The characteristic time $t^*$ is derived from the mean motion of the system and is calculated using $t^* = \sqrt{l^{*3}/(Gm^*)} = 375{,}200$ seconds, where $G$ is the universal gravitational constant. Using these scales, all distances, times, and masses in the simulation are nondimensionalized. Consequently, state vectors such as position and velocity, as well as control inputs, are expressed in these nondimensional units. This nondimensionalization simplifies the CR3BP equations of motion and renders them independent of the physical units of the Earth–Moon system, enabling more general analysis and interpretation.

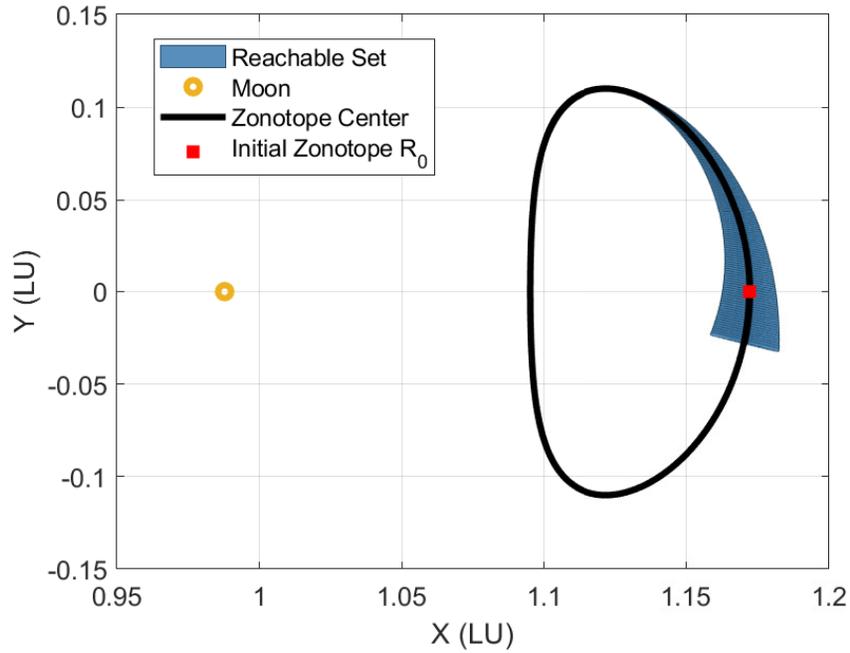

**Figure 2**: Position reachable set for L2 Halo orbit (Method 1). Black curve traces zonotope centers.

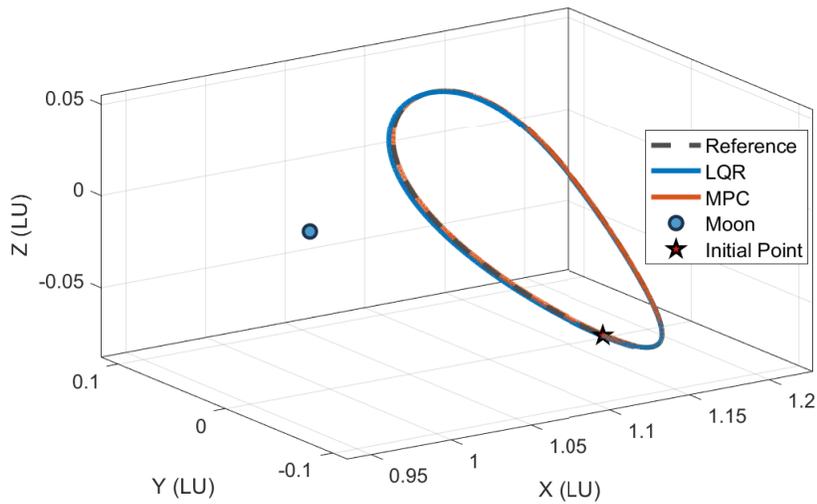

**Figure 3**: Reference tracking performance comparison between MPC and LQR controllers.

Reachability analysis is performed for a low-thrust spacecraft operating in a Halo orbit around the L2 point of the Earth–Moon CR3BP. The spacecraft parameters are set as $T_{\max} = 0.1$ N, $I_{sp} = 3000$ s, and initial mass $m_0 = 1000$ kg. The initial state, expressed in nondimensionalized coordinates, is adopted from [16] and given by $\boldsymbol{x}_0 = [1.1720; 0; -0.0862; 0; -0.1880; 0]^\top$.

The period of the L2 reference orbit is 346.322857 hours. Figure 2 illustrates the evolution of the position reachable set over the given time horizon, computed using the zonotope-based method (Method 1) described in the previous sections. The initial zonotope is centered at $\boldsymbol{x}_0$, with a small generator matrix to approximate the initial uncertainty. The black trajectory shown corresponds to the zonotope centers over time, providing a visual representation of the reference trajectory. The zonotope center trajectory is then used as a reference for feedback tracking via both MPC and LQR. Figure 3 shows the performance of both controllers. While both methods successfully track the reference trajectory, MPC exhibits both lower tracking error and lower total control effort, at modest additional computational cost. (Table 2).

Table 3 summarizes the key parameters used for the MPC implementation. The tuning matrices $\boldsymbol{Q}$ and $\boldsymbol{R}$ were selected to penalize large deviations in position and excessive control effort, respectively.

Table 2: Comparison of LQR and MPC Performance

| **Metric** | **LQR** | **MPC** |
| --- | --- | --- |
| Total Computation Time | 291.8010 s | 305.9837 s |
| Average Time per Iteration | 0.0167 s | 0.0175 s |
| Total Control Effort $\sum_k \|\boldsymbol{u}_k\|$ | 13.579 | 1.385 |
| Total Tracking Error $\sum_k \|\boldsymbol{x}_k - \boldsymbol{x}_k^{\text{ref}}\|$ | 28.7529 | 17.2308 |

Table 3: MPC parameters used in the simulation.

| **Parameter** | **Value** |
| --- | --- |
| Prediction Horizon ($N_p$) | 10 |
| Control Horizon ($N_c$) | 5 |
| Maximum Thrust ($T_{\max}$) | 0.1 N |
| Initial Mass ($m_0$) | 1000 kg |
| Specific Impulse ($I_{sp}$) | 3000 s |
| State Weight Matrix ($\boldsymbol{Q}$) | $\text{diag}([1000, 1000, 1000, 100, 100, 100])$ |
| Control Weight Matrix ($\boldsymbol{R}$) | $\text{diag}([1, 1, 1])$ |

**L2 9:2 Near Rectilinear Halo Orbit: Method Comparison**

The reachable set for a low-thrust spacecraft operating in a L2 Halo orbit is analyzed. The spacecraft is modeled with $T_{\max} = 0.1$ N, $I_{sp} = 3000$ s, and $m_0 = 1000$ kg. initial state vector in nondimensionalized CR3BP coordinates is given by $\boldsymbol{x}_0 = [1.1720, 0, -0.0862, 0, -0.1880, 0]^\top$. Figure 4 presents a direct comparison of the reachable sets computed over a 76 h time horizon by the direct Taylor expansion method and the SDC parameterization. The left panel shows the result from the method 1, which relies on a direct Taylor expansion of the nonlinear dynamics. As the trajectory approaches the Moon, the Lagrange remainder term grows significantly due to the strong gravitational nonlinearities, causing a rapid expansion of the computed zonotopes. In contrast, the right panel shows that the SDC method provides a much tighter and more stable reachable set in this same region. By embedding the nonlinearities within the state-dependent matrix, the SDC parameterization better contains the wrapping effect, demonstrating superior performance and yielding a less conservative approximation. This stability is what makes long-term propagation feasible, enabling the computation of the reachable set over a full orbit, as shown in Figure 6, a task that is intractable with Method 1 due to its unbounded error growth near the Moon.

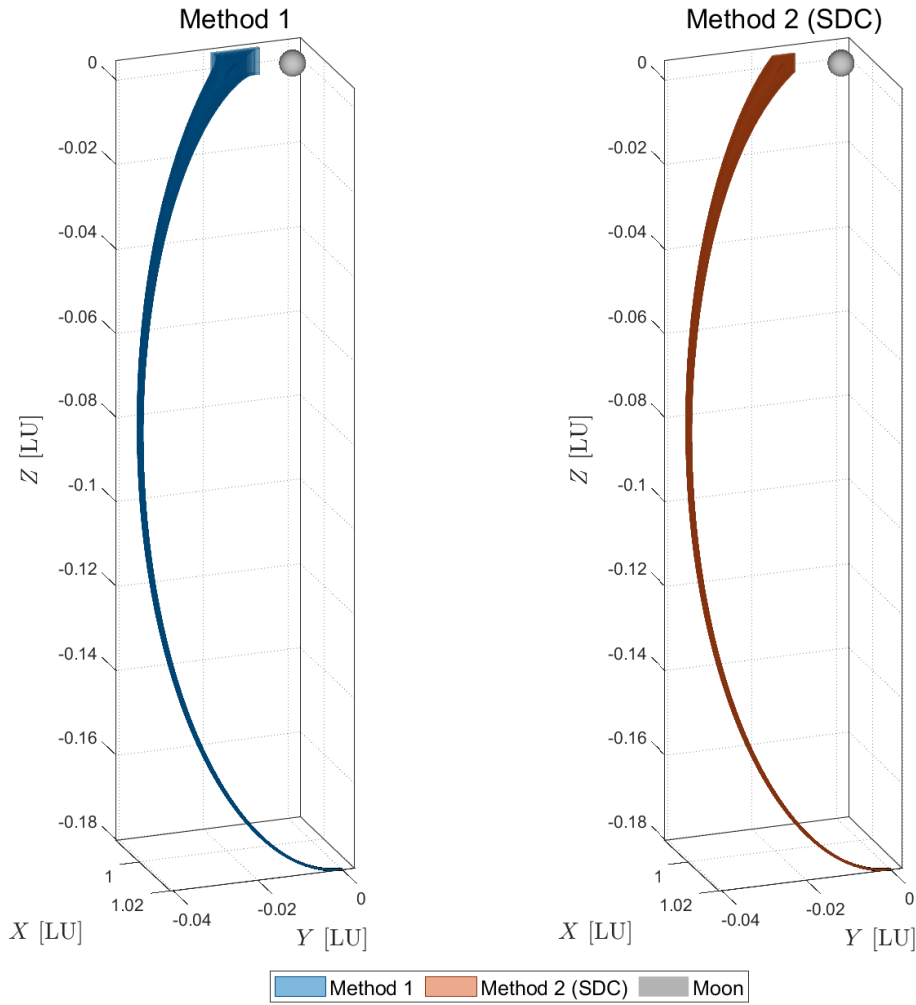

**Figure 4**: 3D Reachable Set Comparison.

This difference in performance is quantified in Figure 5, which plots the logarithm of the reachable set's position-space volume over time. The volume of the set generated by Method 1 exhibits a steep, exponential-like growth, confirming the rapid loss of precision observed visually. Conversely, the SDC method's volume grows at a much slower, more controlled rate.

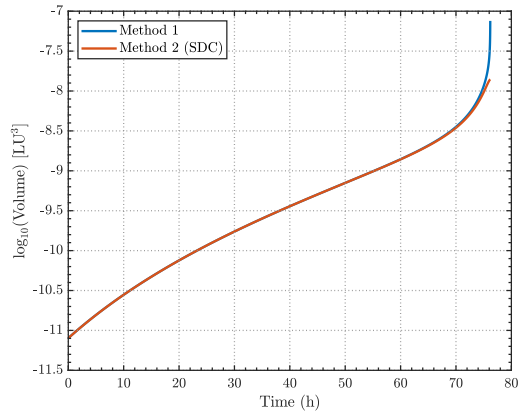

Figure 5: Comparison of Reachable Set Volume (Conservativeness).

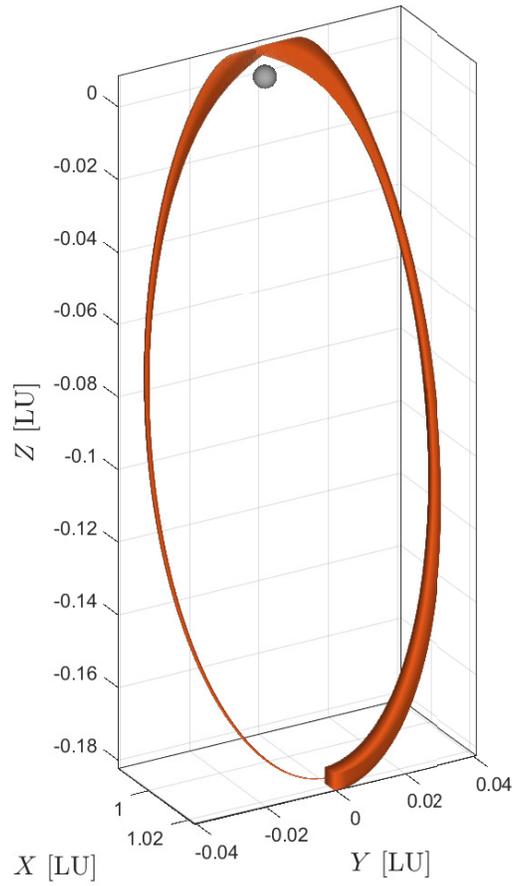

Figure 6: 3D Reachable Set Evolution (SDC Method).

**Station-Keeping in the Near Rectilinear Halo Orbit**

A simulation with L2 reference orbit of period $t_f = 157$ h was performed with the online SDC reachable set propagation integrated with a model predictive controller. The controller uses a prediction horizon of $N = 15$ steps. The same initial state vector is used as in the above results, $\boldsymbol{x}_0 = [1.1720, 0, -0.0862, 0, -0.1880, 0]^\top$.

To ensure robustness and constraint satisfaction, the MPC formulation incorporates reachable sets computed using the SDC method. At each apolune, the current nominal state is used to propagate a tube of forward reachable sets over a short horizon. This tube, denoted $\mathcal{R}_{\text{sdc}}(t)$, characterizes all admissible trajectories under bounded disturbances and control inputs. The controller monitors whether the actual trajectory exits this tube; if so, a station-keeping maneuver is triggered. Upon activation, the MPC optimizes a control sequence that ensures the trajectory re-enters the interior of the reachable set and remains within it. This tube-based constraint enforcement avoids unnecessary firings. Once the spacecraft is safely back inside the predicted tube, MPC is deactivated until the next event trigger.

Figure 7 overlays the reference orbit (red dashed) and the controlled spacecraft track (blue) in the three-dimensional planes. Yellow markers denote time instants when the MPC was *enabled*.

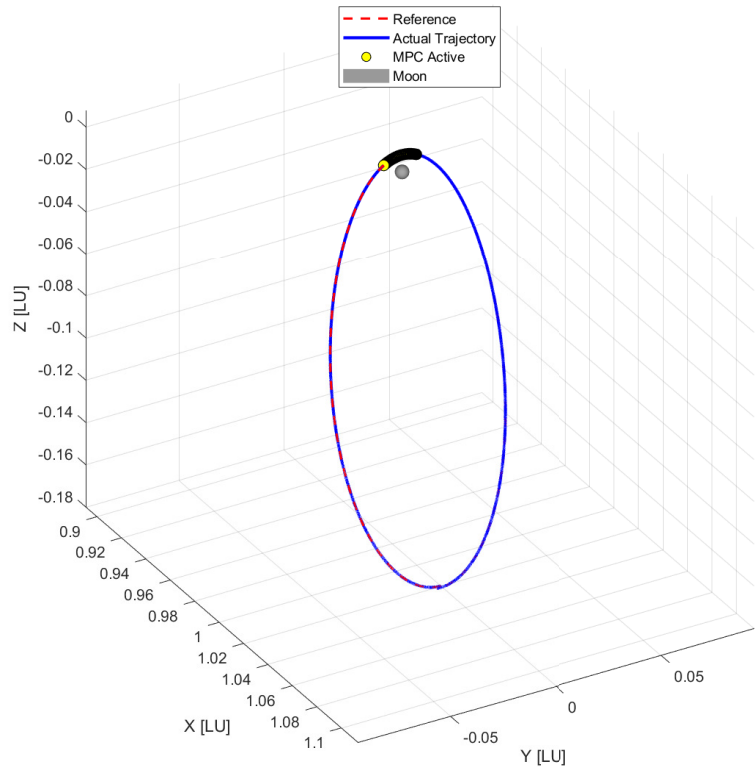

**Figure 7**: Station-keeping performance on 9:2 NRHO. Yellow circles indicate MPC activation events.

The control history is plotted in Figure 8. Each burn is a short, smooth pulse well within the imposed bound. Integrating the dimensional acceleration over the entire simulation gives a cumulative

$$\Delta V_{\text{tot}} = 0.0061 \text{ m/s},$$

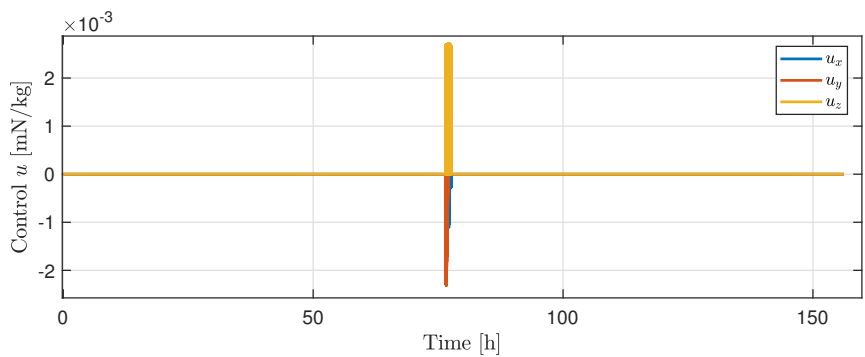

**Figure 8**: Control Input.

The pale blue boxes in Figure 7 visualise the axis-aligned bounding box of the current SDC reachable set. After the initial corrections the spacecraft re-entered the shrunk tube before perilune, deactivating the MPC. This *hysteresis* prevents chatter and demonstrates that updating the tube once per orbit is sufficient for robust containment over multi-revolution horizons.

All disturbances in the simulation are explicitly modeled as bounded sets using zonotopes. The initial state uncertainty is captured by a zonotope with generator matrices $\mathbf{G}_{\text{pos}} = 1 \times 10^{-4}\,\mathbf{I}_3$ for position and $\mathbf{G}_{\text{vel}} = 5 \times 10^{-4}\,\mathbf{I}_3$ for velocity, where $\mathbf{I}_3$ is the $3 \times 3$ identity matrix. Additionally, the process disturbance acting at each time step is represented by a zonotope generator $\mathbf{W}_{\text{gen}} = 1 \times 10^{-7}\,\mathbf{I}_3$, which bounds the unmodeled accelerations and external perturbations. All values are specified in nondimensional CR3BP units. This approach ensures that the reachable sets computed for the spacecraft's evolution accurately account for both the initial uncertainty and ongoing disturbances, maintaining rigorous safety and feasibility guarantees in the station-keeping strategy.

As the spacecraft traverses the orbit, these disturbances may accumulate, causing the actual trajectory and the predicted reachable set to gradually diverge from the nominal reference path. Whenever the current spacecraft state exits the computed SDC tube, the station-keeping logic detects this event and activates MPC to trigger a corrective burn. This control action re-injects the spacecraft state inside the safe tube. Since the timing and direction of the disturbances are arbitrary, station-keeping events can occur at any point along the NRHO. The integration of zonotope-based reachability and feedback control thus ensures robust, set-theoretic constraint satisfaction and maintains the spacecraft within operational bounds despite persistent, bounded uncertainties.

**Remark.** *This work showcased a set-based reachability pipeline, formulation, computation, and use via two methods. Beyond theory, pre-computed reachable sets speed mission design, give instant feasibility bounds, and act as safety envelopes; here we embedded them in MPC for station-keeping. Results were obtained in the idealised CR3BP; therefore the per-orbit $\Delta V$ and controller-activation counts reported here will shift when high-fidelity ephemerides and perturbations are introduced. The same framework, however, is ready to support on-board replanning, contingency burns, and rendezvous once those higher-order models are integrated and validated over multi-year horizons.*

**CONCLUSION**

Two practical methods for nonlinear reachable-set computation have been presented and validated. The first relies on a direct high-order Taylor approximation of the dynamics; the second employs an SDC (state-dependent coefficient) parameterisation that sharply bounds the nonlinear remainder and maintains tight set growth even near strong gravitational perturbations. The techniques were applied to (i) an Earth-to-Mars transfer under two-body dynamics and (ii) a cislunar L2 Halo orbit governed by CR3BP equations, where the SDC method proved both more accurate and numerically stable. In the CR3BP case, the computed reachable-set tube was successfully combined with a model-predictive controller to achieve fuel-efficient, event-triggered station-keeping. Future work will extend the SDC framework to near-rectilinear Halo orbits, incorporate high-fidelity disturbance models (solar radiation pressure, luni-solar perturbations), and deliver a complete MPC guidance architecture for long-duration, low-thrust cislunar missions.